\documentclass[12pt,a4paper]{article}
\usepackage{amsmath}
\usepackage{bm}

\title{
Geometrical interpretation\\ of Hall conductivity in metals}

\author{
Osamu Narikiyo\thanks{Department of Physics, Kyushu University, 
744 Motooka, Fukuoka 819-0395, Japan}
}

\begin{document}
\maketitle

{\it The weak-field Hall conductivity in metals is interpreted 
in terms of the curvature of the Fermi surface in the main part. 
In the appendix 
the orbital magnetic-susceptibility and the magneto-conductivity 
in metals are discussed focusing on Peierls' area factor.}
\\
\\

Haldane~\cite{Haldane} aimed to express 
the Hall conductivity in metals under weak magnetic field 
in terms of the curvature of the Fermi surface. 
However, the geometrical formula obtained by him is incorrect 
and we showed the correct formula in Ref. 2. 
In these works the curvature is derived 
from the directions of the vectors normal to the Fermi surface. 

In our semi-classical approximation 
the cyclotron motion of electrons determines the Hall conductivity. 
However, the above-mentioned derivation 
is not directly related to the cyclotron motion. 
Thus in this paper we will give another derivation 
of the geometrical expression of the Hall conductivity 
on the basis of the tangent vector to the cyclotron orbit in ${\bm k}$-space. 

Since most materials necessary for the derivation are already given in Ref. 2, 
we will borrow them in the following. 
We will use the same notation as in Ref. 2. 
Mathematical facts necessary for the derivation 
can be found in elementary textbooks on differential geometry. 
See, for example, Ref. 3. 

First, we we will discuss the physics. 
Let us confirm the Hall conductivity  
in the semi-classical approximation by the Chambers formula. 
See, for example, Eq. (13.69) in Ref. 4. 
If ${\bm E}=(0,E,0)$ and ${\bm B}=(0,0,B)$, 
the off-diagonal conductivity $\sigma_{xy}$ is given as 
\begin{equation}
\sigma_{xy} = 
e^2 \sum_{\bm k} \tau v_x {\bar v_y} 
\bigg( -\frac{\partial f}{\partial\varepsilon}  \bigg),  
\label{Chambers1}
\end{equation}
where $\bar {\bm v}$ is the average of the velocity 
over the past orbit in $\bm k$-space. 
In the limit of weak magnetic field, $\bar v_y$ is estimated 
to be $v_y(-\tau) = v_y - \dot v_y \tau$ so that we obtain 
\begin{equation}
\sigma_{xy} = 
- e^2 \sum_{\bm k} \tau^2 v_x {\dot v_y} 
\bigg( -\frac{\partial f}{\partial\varepsilon}  \bigg).  
\label{Chambers2}
\end{equation}

In the case of spherical symmetry, 
we evaluate ${\dot v_y}$ by the equation of motion 
$\dot{\bm v} = (e/m) ({\bm v}\times{\bm B})$ in the absence of electric field 
so that ${\dot v_y} = - (eB/m) v_x = \omega_{\rm c} v_x$. 
Then we obtain $\sigma_{xy} = - \omega_{\rm c} \tau \sigma_{xx}$ 
where the DC conductivity in the limit of weak electric field is given by 
\begin{equation}
\sigma_{xx} = 
e^2 \sum_{\bm k} \tau v_x v_x 
\bigg( -\frac{\partial f}{\partial\varepsilon}  \bigg).  
\label{DC}
\end{equation}

In general case, ${\dot v_y}$ is determined by the equation of motion, 
$\dot{\bm v} = e M^{-1} ({\bm v}\times{\bm B})$ in the absence of electric field. 
Since $ {\dot v_y} = e ( M^{-1}_{yx} v_y - M^{-1}_{yy} v_x) B $
in our case of  ${\bm B}=(0,0,B)$, we obtain 
\begin{equation}
\sigma_{xy} = 
e^3 \sum_{\bm k} \tau^2 v_x \big( M^{-1}_{yy} v_x - M^{-1}_{yx} v_y \big) B  
\bigg( -\frac{\partial f}{\partial\varepsilon}  \bigg).  
\label{Hall}
\end{equation}
This expression is better than Eq. (1) in Ref. 2, 
since it is free from the derivative of $\tau$. 
Here the inverse-mass roughly represents the curvature of the Fermi surface. 
In the case of spherical symmetry, $M^{-1}_{yy}=1/m$ and $M^{-1}_{yx}=0$. 
We obtain Eq. (9) in Ref. 2 if we consider $(\sigma_{xy} - \sigma_{yx})/2$. 

Next, we will discuss the geometry. 
Let us consider a cyclotron orbit in $\bm k$-space. 
The orbit is the intersection 
between the Fermi surface and the plane perpendicular to magnetic field. 
See, for example, Fig. 12.6 in Ref. 4. 
We use a notation $ {\bm k} = (k_x, k_y, k_z) \equiv (x, y, z)$. 
We choose a point on the Fermi surface 
and introduce the tangent plane at the point. 
Let us introduce orthogonal coordinates $(x, y, z)$ 
where the point is at the origin $(0, 0, 0)$ 
and the $x$- and $y$- axes are on the plane. 
In the vicinity of the point 
we consider the curvature $\kappa$ 
along a curve on the Fermi surface through the point. 
We choose the $x$-axis 
such that the curve in the $xz$-plane gives the maximum value 
$\kappa_{\rm max}$ of the curvature. 
In this case the curve in the $yz$-plane gives the minimum value $\kappa_{\rm min}$. 
Here 
$ \kappa_{\rm max} = (\dot x \ddot z - \ddot x \dot z) / (\dot x^2 + \dot z^2)^{3/2}$
and 
$ \kappa_{\rm min} = (\dot y \ddot z - \ddot y \dot z) / (\dot y^2 + \dot z^2)^{3/2}$. 
In the evaluation of $\kappa_{\rm max}$ and $\kappa_{\rm min}$ 
we use $ {\dot {\bm k}} = e ({\bm v}\times{B})$ 
and $\dot{\bm v} = e M^{-1} ({\bm v}\times{\bm B})$. 
For $\kappa_{\rm max}$ we take ${\bm B}=(0,B,0)$ and obtain  
$ \kappa_{\rm max} = 2 h_y / v^2 $. 
For $\kappa_{\rm min}$ we take ${\bm B}=(B,0,0)$ and obtain 
$ \kappa_{\rm min} = 2 h_x / v^2 $. 
Here $h_x$ and $h_y$ are given in Eq. (20) and Eq. (21) in Ref. 2 where 
$2 h_x v =  
\varepsilon_y ( \varepsilon_y  \varepsilon_{zz} - \varepsilon_z  \varepsilon_{zy} ) + 
\varepsilon_z ( \varepsilon_z  \varepsilon_{yy} - \varepsilon_y  \varepsilon_{yz} ) $ 
and 
$2 h_y v =  
\varepsilon_z ( \varepsilon_z  \varepsilon_{xx} - \varepsilon_x  \varepsilon_{xz} ) + 
\varepsilon_x ( \varepsilon_x  \varepsilon_{zz} - \varepsilon_z  \varepsilon_{zx} ) $ 
with $\varepsilon_x = \partial \varepsilon / \partial x$, 
$\varepsilon_{xy} = \partial^2 \varepsilon / \partial x \partial y$ and so on. 
It is exploited that $ \dot x^2 + \dot z^2 = (eB)^2 (v_x^2 + v_z^2) = (eB)^2 v^2 $ 
since $v_y=0$ along the curve for $\kappa_{\rm max}$. 
In the same way  $ \dot y^2 + \dot z^2 = (eB)^2 (v_y^2 + v_z^2) = (eB)^2 v^2 $ 
since $v_x=0$ along the curve for $\kappa_{\rm min}$. 
The Hall conductivity $\sigma_{yz}$ is determined by $h_x$ as Eq. (22) in Ref. 2 
and $\sigma_{zx}$ by $h_y$ as Eq. (23) in Ref. 2 where 
$ \sigma_{yz} = e^3 B \int {\rm d}S / (2\pi)^3 h_x \tau^2 $ and 
$ \sigma_{zx} = e^3 B \int {\rm d}S / (2\pi)^3 h_y \tau^2 $. 

Now it has become clear that the integrand of the Hall conductivity 
is proportional to the curvature along the cyclotron orbit. 
In the above discussion for $\kappa_{\rm max}$ and $\kappa_{\rm min}$ 
we have chosen special local coordinates. 
Although the directions of the coordinates are different 
among points on the Fermi surface, 
the value of the mean-curvature $ H = ( \kappa_{\rm max} + \kappa_{\rm min} )/2 $ 
is independent of the coordinates. 
The mean-curvature is the trace of a $3 \times 3$ matrix 
as Eq. (C.5) in Ref. 2 where the curvature in the tangent plane is zero. 
The value of the trace is unchanged 
if we use any coordinates as Eq. (C.2) in Ref. 2. 
Thus in order to obtain the geometrical quantity 
which is independent of the choice of the coordinates 
we have to consider the trace as Eq. (24) in Ref. 2. 

This Short Note and Ref. 2 is the answer to the question raised by Prof. Hirofumi Wada. 
The author greatly thanks him for advices and encouragements.

\newpage

\noindent
{\Large\bf Appendix: Peierls' area factor}

\vspace{3mm}

In this appendix 
the orbital magnetic-susceptibility and the magneto-conductivity 
in metals are discussed focusing on Peierls' area factor. 

The orbital magnetic-susceptibility $\chi$ 
for nearly-free electrons in metals is given 
by integrating 
\begin{equation}
R = \varepsilon_{xx} \varepsilon_{yy} 
  - (\varepsilon_{xy})^2,
\label{area}
\end{equation}
over the Fermi surface. 
Such a formula was obtained by Peierls 
and written in his textbook~\cite{Peierls} as (7.44). 
Here we review how $R$ appears in $\chi$ 
and explain that $R$ also appears 
in the magneto-conductivity $\Delta \sigma$. 
This appendix is an extension of Ref. 2 and Ref. 3 
and we employ the same framework and notation therein. 

First, we introduce the Fermi surface in $\bm k$-space. 
The point on the Fermi surface 
$ {\bm k} = (k_x, k_y, k_z) \equiv (x, y, z)$ 
satisfies the condition 
$\varepsilon(x,y,z)=0$ 
where $\varepsilon(x,y,z)$ is the renormalized band-energy of electrons. 
$\varepsilon_x$ stands for $\partial \varepsilon / \partial x$ 
and $\varepsilon_{xy}$ stands for 
$\partial^2 \varepsilon / \partial x \partial y$ and so on. 
We need such derivatives around the Fermi surface 
when we calculate $\chi$ and $\Delta \sigma$. 

Now, the definition of $R$ is fixed. 
$R$ is interpreted as the Gaussian curvature 
in a classic textbook~\cite{Wilson} and 
in some recent papers~\cite{FOF,RPFM}. 
However, it is not. 
The Gaussian curvature of the Fermi surface is given by 
$ ( w_{xx}w_{yy} - w_{xy}^2 ) / ( 1 + w_x^2 + w_y^2 )^2 $
when $z=w(x,y)$. 

Then, we will think over the interpretation of $R$. 
In our semi-classical treatment~\cite{Narikiyo1,Narikiyo2} 
of the weak magnetic field 
the effect of the magnetic field ${\bm B}$ is taken into account 
by the Peierls-Schwinger factor~\cite{Edelstein}
\begin{equation}
F({\bm r}', {\bm r}) 
= \exp \Bigg[ i e  
  \int_{\bm r}^{{\bm r}'} d{\bm s} \cdot {\bm A}({\bm s}) \Bigg]
= \exp \Bigg[ i e 
{1 \over 2} {\bm B} \cdot ({\bm r}\times{\bm r}') \Bigg],
\label{P-S}
\end{equation}
where we have employed the symmetric gauge 
${\bm A} = {1 \over 2} {\bm B}\times{\bm r}$ 
and ${\bm B}$ is a constant vector. 
Namely, the effect is represented by the magnetic flux 
penetrating the triangle specified by the vectors $\bm r$ and ${\bm r}'$ 
whose area is ${1 \over 2}|{\bm r}\times{\bm r}'|$. 

$\chi$ is determined by the energy change in the order of  
$[{\bm B} \cdot ({\bm r}\times{\bm r}')]^2$. 
In $\bm k$-space $\bm r$ acts as 
$-i(\partial/\partial x, \partial/\partial y, \partial/\partial z)$. 
The energy change is given by the integral of the action of 
$({\bm r}\times{\bm r}')^2$ on $\varepsilon$. 
Via integration by parts we obtain the integral containing $R$~\cite{WU}. 
Thus $R$ stems from the area in $\bm r$-space 
and we call $R$ the area factor. 

Next, we will show that 
$R$ also appears in the magneto-conductivity $\Delta \sigma$. 
In the following we fix the constant electric field 
${\bm E}=(E,0,0)$ in $x$-direction 
and the constant magnetic field ${\bm B}=(0,0,B)$ in $z$-direction. 
In our semi-classical treatment 
the solution of the Boltzmann equation gives~\cite{Ziman}  
\begin{equation}
\Delta \sigma = 
e^4 B^2 \sum_{\bm k} \tau v_x 
\Bigg[
\bigg( v_x {\partial \over \partial y} 
     - v_y {\partial \over \partial x} \bigg) \tau \Bigg]^2
v_x
\bigg( -\frac{\partial f}{\partial\varepsilon} \bigg).  
\label{Boltzmann}
\end{equation}
Here $v_x \partial/\partial y - v_y \partial/\partial x$ 
can be interpreted~\cite{Enz} 
as the rotation or angular-momentum operator around $z$-axis 
in $\bm k$-space. 
It is evident in the case of spherical symmetry, 
because ${\bm v} = {1 \over m} (x,y,z)$. 
Then $\tau(v_x \partial/\partial y - v_y \partial/\partial x)$ 
is proportional to the area in $\bm k$-space. 

Since we are interested in the dominant contributions in $\tau$, 
we neglect the derivative of $\tau$ in (\ref{Boltzmann}). 
In the study of the Hall conductivity 
for nearly-free electrons~\cite{FEW} and Fermi liquids~\cite{KY} 
only the dominant contributions proportional to $\tau^2$ 
are considered. 
In the same manner 
we only consider the dominant $\tau^3$ contributions 
in (\ref{Boltzmann}) and obtain 
\begin{equation}
\Delta \sigma = 
- e^4 B^2 \sum_{\bm k} \tau^3 v_x R v_x
\bigg( -\frac{\partial f}{\partial\varepsilon} \bigg).  
\label{mc}
\end{equation}
By comparison with the conductivity for ${\bm B}=0$ 
\begin{equation}
\sigma = 
e^2 \sum_{\bm k} \tau v_x v_x
\bigg( -\frac{\partial f}{\partial\varepsilon} \bigg),  
\label{dc}
\end{equation}
we see that the effect of the magnetic field 
is represented by the area factor as $(eB\tau)^2 R$. 

In a specific case of spherical symmetry 
$(eB\tau)^2 R$ reduces to $(\omega_{\rm c}\tau)^2$ 
where $\omega_{\rm c} = -eB/m$ 
with $\varepsilon_{xx}=\varepsilon_{yy}=1/m$ and 
$\varepsilon_{xy}=0$. 
Thus we can say that 
$\omega_{\rm c}\tau$ represents the magnetic flux. 

In general cases we can define the cyclotron effective-mass $m^*$ 
by $(m^*)^{-2} = R$. See (12.65) and (28.7) in Ref. 13. 
Since the cyclotron effective-mass is related to an area in $\bm k$-space, 
$R$ is also related to the area. 

We can also reach (\ref{mc}) starting from the Chambers formula. 
In the semi-classical approximation 
the Chambers formula for the diagonal conductivity becomes 
\begin{equation}
\sigma_{xx} = 
e^2 \sum_{\bm k} \tau v_x {\bar v_x} 
\bigg( -\frac{\partial f}{\partial\varepsilon}  \bigg),  
\label{Chambers1}
\end{equation}
where $\bar {\bm v}$ is the average of the velocity 
over the past orbit in $\bm k$-space. 
See, for example, (13.69) in Ref. 13. 
In the limit of weak magnetic field, 
$\bar v_x$ is estimated~\cite{Kittel} to be 
$\bar v_x = v_x - \dot v_x \tau + \ddot v_x \tau^2 + \cdot\cdot\cdot$ 
neglecting the derivatives of $\tau$ so that we obtain 
\begin{equation}
\Delta \sigma = 
e^2 \sum_{\bm k} \tau^3 v_x {\ddot v_x} 
\bigg( -\frac{\partial f}{\partial\varepsilon}  \bigg).  
\label{Chambers2}
\end{equation}
${\ddot v_x}$ is determined by the equation of motion 
$\dot{\bm v} = e M^{-1} ({\bm v}\times{\bm B})$ 
in the absence of electric field. 
Here $M^{-1}_{ij} = \varepsilon_{ij}$ 
with $i,j = x,y,z$. 
See Ref. 3. 
Thus we obtain (\ref{mc}). 

\vspace{6mm}

\noindent
{\Large\bf Appendix: Hall conductivity is proportional to diamagnetic susceptibility}

\vspace{3mm}

In our old paper (J. Phys. Soc. Jpn. 93 (1993) 1812) 
we pointed out that the Hall conductivity is proportional to the diamagnetic susceptibility. 
But the explanation is insufficient. 
Here we will try to make a precise explanation.  

The Hall conductivity $\sigma_{xy}$ is proportional to 
$ \langle v_x G G v_y \rangle$ 
where $G$ is the propagator in the magnetic field $H$. 
$H$ is proportional to $ q_x A_y - q_y A_x $ (See arXiv:1203.0127v2). 
The $q_x$-linear contribution from $G$ is $G_0 v_x q_x G_0$ 
where $G_0$ is the propagator in the absence of the magnetic field. 
Similarly, the $A_y$-linear contribution from $G$ is $G_0 e v_y A_y G_0$. 
Consequently, $\sigma_{xy}$ is proportional to 
$ \langle v_x G_0 v_y G_0 v_x G_0 v_y G_0\rangle$. 

On the other hand, the diamagnetic susceptibility $\chi$ is a $H^2$ contribution 
in the thermodynamic potential. 
It is given by the closed fermion loop proportional to  $ (q_x A_y - q_y A_x)^2 $ 
so that $\chi$ is proportional to $ \langle v_x G_0 v_y G_0 v_x G_0 v_y G_0\rangle$.

\end{document}